\documentclass[aps,prl,showpacs,floatfix,twocolumn,longbibliography]{revtex4-1}
\usepackage{mathrsfs}
\usepackage[figuresright]{rotating}
\usepackage{amsmath}
\usepackage{amssymb}
\usepackage{siunitx}
\usepackage{graphicx}% Include figure files
\usepackage{color}
\usepackage{dcolumn}% Align table columns on decimal point
\usepackage{bm}% bold math
\usepackage[breaklinks=true,colorlinks=true,linkcolor=blue,urlcolor=blue,citecolor=blue]{hyperref}
\usepackage{ragged2e}
\UseRawInputEncoding   % This is to solve the problem of "Invalid UTF-8 byte "A1"
\usepackage{longtable}
\usepackage{multirow}
\usepackage{setspace}
\usepackage{soul}

\makeatletter

\newcommand{\Rmnum}[1]{\expandafter\@slowromancap\romannumeral #1@}
\newcommand{\supplementtableofcontents}{%
  \par
  \vspace{1em}
  \begin{center}
    \bfseries CONTENTS
  \end{center}
  \vspace{-0.5em}
  \@starttoc{smtoc}%
  \par
}
\newcommand{\SMsection}[1]{%
  \section{#1}%
  \addcontentsline{smtoc}{section}{%
    \protect\numberline{\thesection}#1%
  }%
}
\makeatother

\begin{document}

\title{$^{139}$La nuclear quadrupole resonance studies of pressurized La$_4$Ni$_3$O$_{10}$}

\author{Meng Zhang$^{1}$}
\author{Zhuo Wang$^{1}$}
\author{Yantao Cao$^{2,3}$} % caoyt@iphy.ac.cn
\author{Yang Yuan$^{1}$}
\author{Kangjian Luo$^{1}$}
\author{Shanxiang Gao$^{1}$}
\author{Hanjie Guo$^{3}$}
\email[]{hjguo@sslab.org.cn}
\author{Yongkang Luo$^{1}$}
\email[]{mpzslyk@gmail.com}
\address{$^1$Wuhan National High Magnetic Field Center and School of Physics, Huazhong University of Science and Technology, Wuhan 430074, China;}
\address{$^2$Institute of Physics, Chinese Academy of Sciences, Beijing 100190, China;}
\address{$^3$Songshan Lake Materials Laboratory, Dongguan, Guangdong 523808, China.}

\date{\today}

\begin{abstract}

Density-wave (DW) orders are considered as competing orders to unconventional superconductivity and are commonly seen in a variety of superconductors including but not limited to the recently discovered Ruddlesden-Popper-phase nickelates. By utilizing $^{139}$La nuclear quadrupole resonance, we systematically investigate into the nature of DW orders and their evolution under pressure in La$_4$Ni$_3$O$_{10}$. Spin and charge DW orders are found to be intertwined in this material, which is in stark contrast to those in La$_3$Ni$_2$O$_7$. Short-range DW orders are observed near 150 K, well above the development of long-range DW orders at around 139 K. Upon applying a hydrostatic pressure of 2.3 GPa, the transition temperatures of the short-range and long-range orders decrease at rates of 1 K/GPa and 10 K/GPa, respectively. Our results thus affirm that both spin density wave and charge density wave as competing orders with the superconducting state in La$_4$Ni$_3$O$_{10}$, and provide new insights into the interplay between DW orders and unconventional superconductivity.

\end{abstract}

%\pacs{75.75.-c, 72.55.+s, 75.30.Kz, 75.10.-b}
%75.75.-c   Magnetic properties of nanostructures
%72.55.+s 	Magnetoacoustic effects
%75.30.Kz 	Magnetic phase boundaries (including magnetic transitions, metamagnetism, etc.)
%75.10.-b 	General theory and models of magnetic ordering

\maketitle

\section{\Rmnum{1}. Introduction}
The recent discovery of pressure-induced superconductivity (SC) with a critical temperature ($T_c$) exceeding 80 K in bilayer Ruddlesden-Popper (RP) structured nickelates $\mathrm{La}_3\mathrm{Ni}_2\mathrm{O}_7$ \cite{La3Ni2O7-Nature2023,327poly-PRX2024,La2PrNi2O7-Nature2024,La3Ni2O7-100Gpa-NSR2025,96KLa3Ni2O7-Nature2026} establishes a novel correlated high-$T_c$ superconductor family alongside cuprates \cite{Bednorz-LBCO} and iron-based \cite{Hosono-LaOFFeAs} superconductors. Extensive studies have been motivated to look into the interplay between competing orders and unconventional SC in these layered systems \cite{LiuZ-La327Correlation_NC2024,LiuZ-RPCorrelation_PRB2025,327-NC2025,327-arxiv2026-SDW,327-arxiv2026-Planckian,327-PRR2026-Oxygen-isotope}. Stoichiometric RP-type nickelates adopt the general chemical formula $Ln_{n+1}\mathrm{Ni}_n\mathrm{O}_{3n+1}$ ($Ln = \text{Lanthanides}$) \cite{Lan+1NinO3n+1-2000}; besides the bilayer $\mathrm{La}_3\mathrm{Ni}_2\mathrm{O}_7$ ($n=2$), the trilayer $\mathrm{La}_4\mathrm{Ni}_3\mathrm{O}_{10}$ ($n=3$) was also reported to exhibit signatures of SC under high pressure \cite{4310poly-CPL2024,4310-PRX2025,4310-PRX2026}, reaching a maximum $T_c\approx 30~\mathrm{K}$ \cite{4310-Nature2024}. Systematic comparisons across these structural variants are essential to elucidate the microscopic origin of SC.

At ambient pressure, both La$_3$Ni$_2$O$_7$ and La$_4$Ni$_3$O$_{10}$ display some kind of density-wave (DW) orders at low temperatures \cite{327and4310-PRB2001,327and4310-SciBull2025,327AND4310-NP2017}. Under pressure, they undergo a structural phase transition, and meanwhile, the DW orders are assumed to be suppressed before SC state emerges \cite{327-NC2025,4310-NC2025}. In La$_3$Ni$_2$O$_7$, two kinds of DW orders were observed, a spin-density-wave (SDW) transition at $\sim 150$ K \cite{327NQR-CPL2025,327NMR-Sci2025,327-NatCommun2024}, followed by another DW transition at $\sim 130$ K whose nature remains unclear yet \cite{327split-NP2025,327NMR-Sci2025,327-NatCommun2024}. These two DW orders seem to be decoupled, and evolve differently under pressure \cite{327split-NP2025,327-2025PRB-coexist,327NMR-Sci2025}. The situation in La$_4$Ni$_3$O$_{10}$, however, appears more elusive in that the SDW and charge-density-wave (CDW) transitions were suggested to take place simultaneously \cite{4310intertwined-NC2020,4310NQR-PRB2026,4310uSR-PRR2026,4310optical-PRB2025,4310charge-PRB2025,4310L-C-Couple-PRX2026}. A natural question then concerns whether they remain intertwined and how they evolve under pressure. To clarify this issue, \textit{local and microscopic} measurements that can distinguish the spin and charge degrees of freedom are needed.

Nuclear quadrupole resonance (NQR) exploits $I>1/2$ nuclei as local probes \cite{Slichter}, where $I$ is quantum number of nuclear spin. Via hyperfine coupling to both internal magnetic field and electric field gradient (EFG), hopefully, the information about spin and charge orders can be extracted and disentangled by NQR. Here, we report systematic $^{139}$La NQR studies on single crystalline La$_4$Ni$_3$O$_{10}$ at ambient and hydrostatic (2.3 GPa) pressures. Our results unveil the appearance of short-range DW orders prior to the long-range orders. The intertwining of SDW and CDW orders is manifested by a non-vanishing magnetic-quadrupolar coupling term. Under pressure, both the long-range and short-range orders are suppressed, at the rates of 10 K/GPa and 1 K/GPa, respectively. These results are rather different from those in La$_3$Ni$_2$O$_7$, and establish both SDW and CDW as competing orders to SC, offering new insights into the mechanism of SC in RP-phase nickelates.

%Before the emergence of superconductivity, both La$_3$Ni$_2$O$_7$ and La$_4$Ni$_3$O$_{10}$ exhibit structural phase transitions and density-wave (DW) transitions \cite{La3Ni2O7-Nature2023,4310-Nature2024,327and4310-PRB2001,327and4310-SciBull2025,327AND4310-NP2017}. Related studies have established that the density wave phase transition is crucial for superconductivity \cite{96KLa3Ni2O7-Nature2026,327-NC2025,4310-NC2025}. In $\mathrm{La}_3\mathrm{Ni}_2\mathrm{O}_7$, there exist two density wave phase transitions at different temperatures: the high-temperature spin density wave (SDW) transition temperature increases with pressure, while the low-temperature charge density wave (CDW) transition temperature decreases with pressure \cite{327NQR-CPL2025,327NMR-Sci2025,327split-NP2025}. When the pressure reaches the critical point where the superconducting phase appears, CDW is fully suppressed. This behavior closely resembles the competition between charge order and superconductivity observed in cuprates \cite{1Cu-Nature2011,2Cu-NP2012,3Cu-Science2018,4Cu-PRB2014}. In $\mathrm{La}_4\mathrm{Ni}_3\mathrm{O}_{10}$, the situation becomes more complex due to the intertwining of incommensurate charge density waves and spin density waves \cite{4310intertwined-NC2020,4310NQR-PRB2026,4310uSR-PRR2026,4310optical-PRB2025,4310charge-PRB2025,4310L-C-Couple-PRX2026}. Therefore, elucidating the density?wave transitions in $\mathrm{La}_4\mathrm{Ni}_3\mathrm{O}_{10}$ is crucial to uncovering the relationship between density?wave order and superconductivity in RP?phase nickelates.

\begin{figure*}[!ht]
\vspace{7pt}
\hspace{-0pt}
\includegraphics[width=17cm]{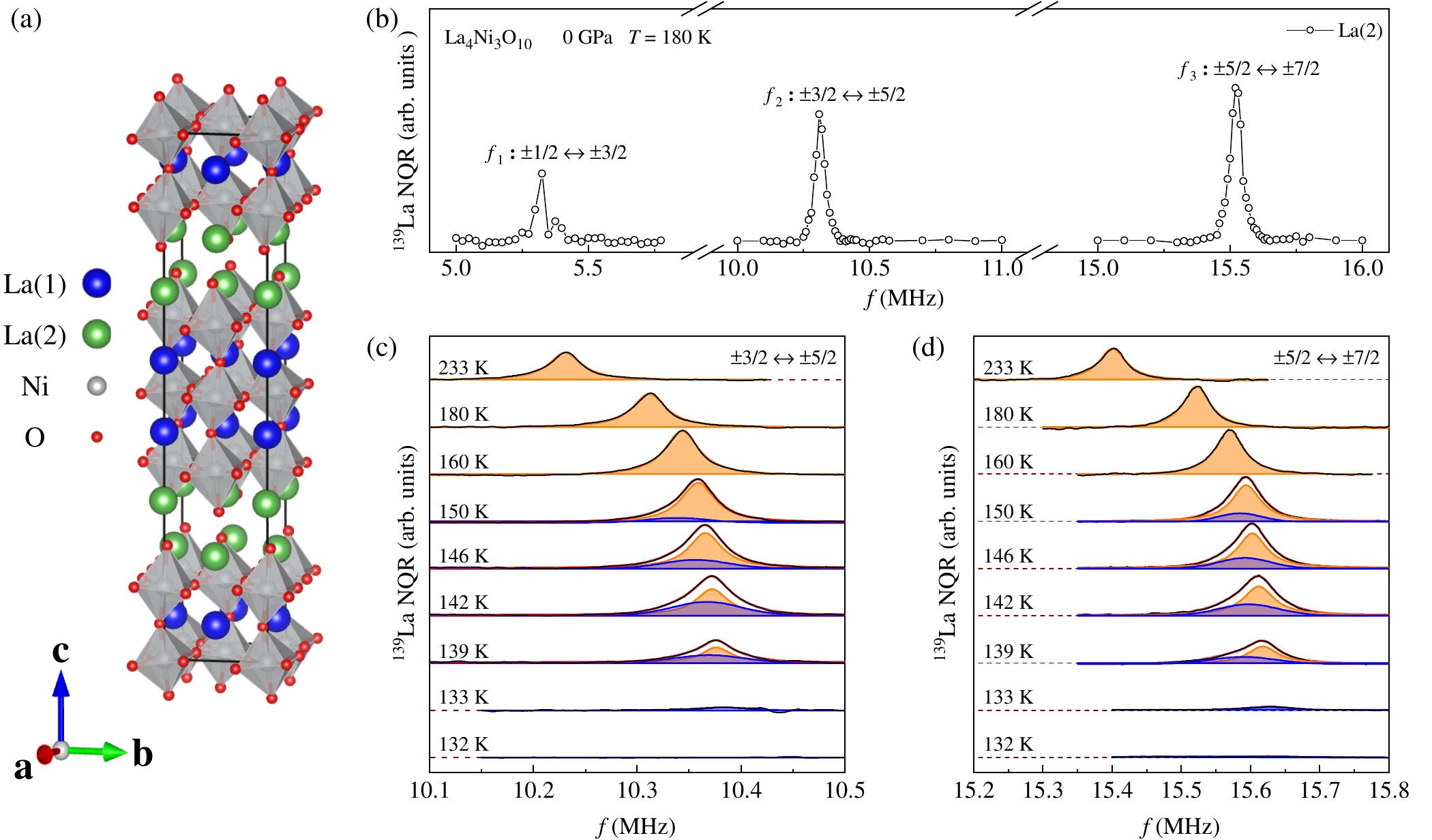}
\vspace*{-5pt}
\caption{(a) Crystal structure of La$_4$Ni$_3$O$_{10}$ showing two distinct La sites. The cell is doubly expanded to show the quasi-tetragonal structure. (b) $^{139}$La(2) NQR spectrum at ambient pressure and 180 K. All three NQR transitions for nuclear spin $I=7/2$ are identified. (c) The $\pm 3/2 \leftrightarrow \pm 5/2$ spectra at selected temperatures. The spectra are verticallly shifted for clarity. Above 150 K, the peak conforms exclusively to a Lorentzian lineshape (orange). Below this temperature, an additional Gaussian contribution (blue) is required. (d) ibid, but for the $\pm 5/2 \leftrightarrow \pm 7/2$ transition.}
\label{Fig1}
\end{figure*}

\section{\Rmnum{2}. Experimental details}

High-quality single crystalline $\mathrm{La}_4\mathrm{Ni}_3\mathrm{O}_{10}$ studied in this work was grown by the high-pressure optical floating-zone technique \cite{4310uSR-PRB2025}. The samples were verified by magnetic susceptibility measurements in a magnetic property measurement system (MPMS, Quantum Design) equipped with a vibrating sample magnetometer (VSM) option, which confirms the density-wave transition near $T_\text{DW}\approx 139$ K, in agreement with literature \cite{4310-Nature2024} (See Fig.~S1 in \textbf{Supplemental Material (SM)} \cite{SM}). Hydrostatic pressure up to $\sim 2.3$ GPa was applied using a piston-cylinder pressure cell (CTF-HHPC60, TOHO HARMONY), and the pressure was determined by monitoring the in-situ $^{63}$Cu NQR frequency of Cu$_2$O mounted in the same coil \cite{Kitagawa-JPSJ2010}. $^{139}$La NQR measurements for temperatures between 80-300 K were carried out using a custom-built liquid-nitrogen measurement system. $^{139}$La NQR spectra were recorded in a stepped frequency-sweep method, while the spin-lattice relaxation time ($T_1$) was obtained by fitting the recovery curve of the $\pm 5/2 \leftrightarrow \pm 7/2$ transition to the stretched formula
\begin{eqnarray}
\begin{aligned}
M(t)=&M(\infty) \{1-2 F [\frac{2574}{12012}\exp(-(\frac{3t}{T_1})^b)\\
&+\frac{7800}{12012}\exp(-(\frac{10t}{T_1})^b)+\frac{1638}{12012}\exp(-(\frac{21t}{T_1})^b)]\},
\end{aligned}
\label{Eq1}
\end{eqnarray}
where $M(\infty)$, $F$, $T_1$, $b$ are fitting parameters. When the stretching exponent $b=1$, Eq.~(\ref{Eq1}) reduces to the standard fitting formula.

\section{\Rmnum{3}. Results and Discussion}

\subsection{A. $^{139}$La NQR at $p=0$}

\begin{figure*}[!ht]
\vspace*{5pt}
\hspace*{-0pt}
\includegraphics[width=16cm]{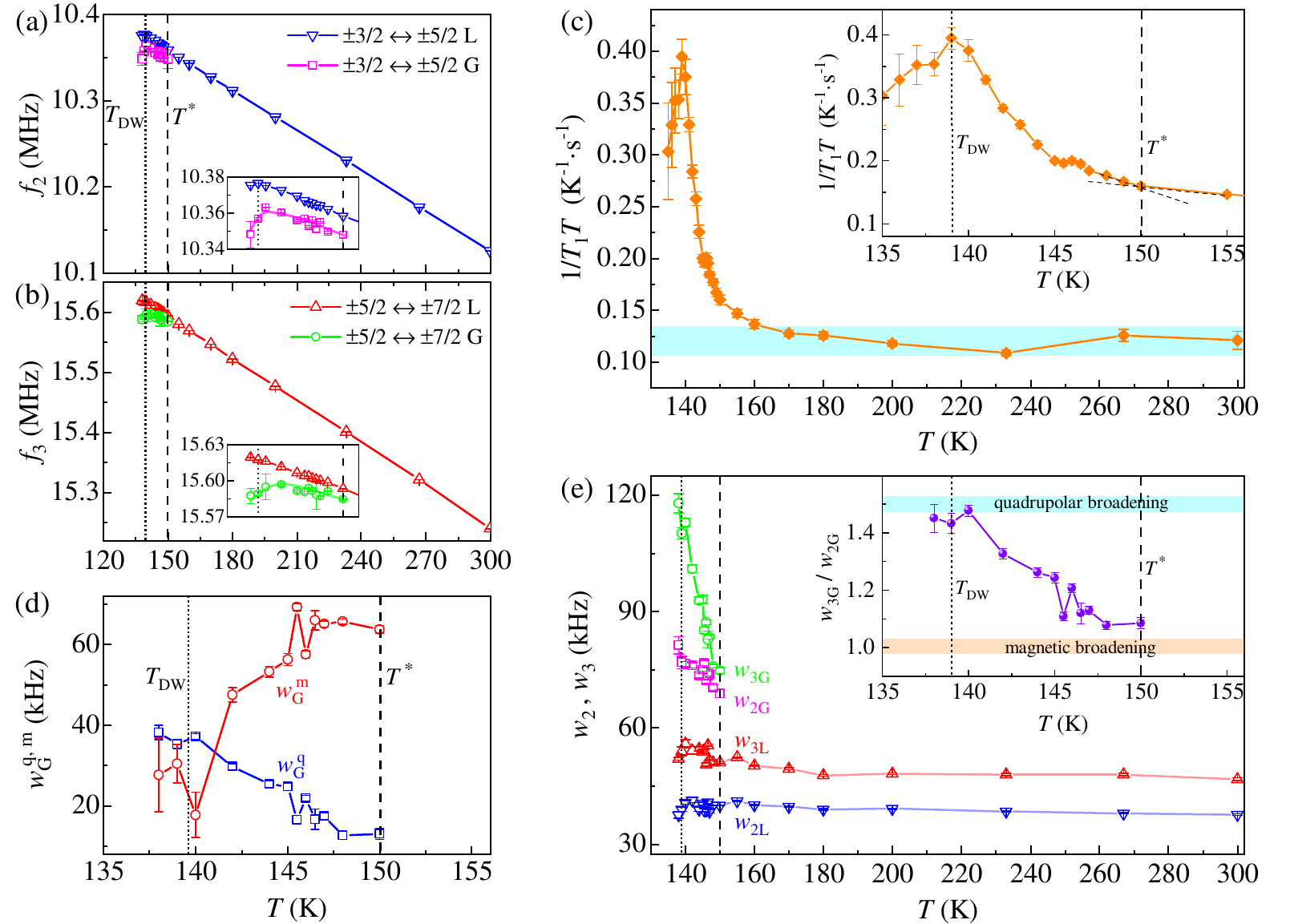}
\vspace*{-3pt}
\caption{(a) and (b) show the temperature dependencies of the peak frequencies
associated with the disordered (Lorentzian-fit) and ordered (Gaussian-fit) states for the $\pm 3/2 \leftrightarrow \pm 5/2 $ and $\pm 5/2 \leftrightarrow \pm 7/2 $ transitions, respectively. The inset provides an enlarged plot near the DW transitions. (c) Temperature dependence of $1/T_1T$ at La(2) site. The inset shows a zoom-in view, as well as the definition of $T_\text{DW}$ and $T^*$. (d) Magnetic ($w^\text{m}_\text{G}$) and quadrupole ($w^\text{q}_\text{G}$) contributions extracted from the linewidth analysis of the ordered state. (e) The separated linewidth contributions to the $f_2$ and $f_3$ peaks. The inset plots the ratio $w_\text{3G}/w_\text{2G}$.}
\label{Fig2}
\end{figure*}

At ambient pressure, $\mathrm{La}_4\mathrm{Ni}_3\mathrm{O}_{10}$ crystallizes in a monoclinic  P2$_1/\text{a}$ structure \cite{4310-Nature2024,4310-AM2025,TTM-arxiv2025,4310transport-SCMA2024}. A doubly expanded cell can be viewed as quasi-tetragonal, as illustrated in Fig.~\ref{Fig1}(a). It contains two crystallographically in-equivalent La sites: La(1) -- the one sits within the NiO$_6$ double layers, and La(2) -- the one resides in the La-O fluorite-type layers outside the NiO$_6$ bilayers. Owing to the much lower NQR frequency and weaker signal intensity of the La(1) site \cite{327and4310-JPSJ2024}, we here only focus on the La(2) site as a local probe. Figure \ref{Fig1}(b) shows the full $^{139}$La(2) NQR spectrum measured at 180 K at ambient pressure. No additional peaks arising from other La$_{n+1}$Ni$_n$O$_{3n+1}$ phases can be identified \cite{327NQR-CPL2025,327NMR-Sci2025,4310NMR-NC2026}.

The nuclear quadrupole Hamiltonian is expressed as
\begin{equation}
    \mathcal{H}_{Q}=\frac{e Q V_{z z}}{4 I(2 I-1)}\left[3 \hat{I}_{z}^{2}-\hat{\mathbf{I}}^{2}+\eta\left(\hat{I}_{x}^{2}-\hat{I}_{y}^{2}\right)\right],
    \label{Eq3}
\end{equation}
where $\hat{\mathbf{I}}=(\hat{I}_{x},\hat{I}_{y},\hat{I}_{z})$ is the nuclear spin operator, $Q$ is nuclear quadrupole moment, and $\eta\equiv (V_{xx}-V_{yy})/V_{zz}$ is the asymmetry parameter with $V_{xx}$, $V_{yy}$ and $V_{zz}$ being the components of the EFG tensor. For $I=7/2$ nuclear spin, three NQR peaks are expected arising from the transitions $\pm 1/2 \leftrightarrow \pm 3/2 $, $\pm 3/2 \leftrightarrow \pm 5/2 $ and $\pm 5/2 \leftrightarrow \pm 7/2$. For brevity, we hereafter refer to them as $f_1$, $f_2$ and $f_3$, respectively. At 180 K, these values are 5.32, 10.31 and 15.53 MHz, consistent with previous reports \cite{327and4310-JPSJ2024,4310NQR-PRB2026,4310NMR-NC2026}.
In Figs.~\ref{Fig1}(c) and (d), we display
the temperature evolution of the $\pm 3/2 \leftrightarrow \pm 5/2 $ and $\pm 5/2 \leftrightarrow \pm 7/2$ transitions. On the whole, both resonance peaks shift towards right-handed upon cooling, as expected. To elucidate the temperature-dependent peak behavior, we implemented a fitting analysis. While a single Lorentzian peak fits the spectra well in the high-temperature regime ($>150$ K), this turns out to be inadequate as lineshape broadening and asymmetry develop below 150 K. Attributing these anomalies to a precursor effect of the short-range DW transition (whose onset temperature is denoted by $T^*=150$ K), we decomposed the spectra into a disordered-state Lorentzian (L) and an ordered-state Gaussian (G) contribution. It should be mentioned here that we did not observe any peak splitting in our NQR spectra at all temperatures, which implies that the DW orders are incommensurate, and this is totally different from La$_{3}$Ni$_2$O$_7$ where commensurate DW order was detected \cite{327NQR-CPL2025,327split-NP2025,327NC-Gupta2025}. Below $T_\text{DW}$, the intensity of the peak shrinks rapidly, and the NQR signals become hardly distinguishable out of the background noise. This behavior is likely attributable to the magnetic wipe-out effect when approaching the SDW transition. The coexistence of L and G components below 150 K manifests the presence of short-range ordering prior to the long-range DW orders, in agreement with a recent report \cite{4310NMR-NC2026}. The temperature dependence of $f_2$ and $f_3$ are extracted and summarized in Fig.~\ref{Fig2}(a-b). Quasi-linear temperature dependence is found in both $f_{2\text{L}}(T)$ and $f_{3\text{L}}(T)$ above $T_\text{DW}$, in accordance with the Bayer-Kushida relation \cite{NSS2020}, whereas no anomaly is discernible across $T^*$. The $f_{2\text{G}}$ and $f_{3\text{G}}$, in contrast, exhibit small changes about $T^*$ with respect to the L counterparts, cf the insets to Fig.~\ref{Fig2}(a-b). (The subscripts ``L" and ``G" denote the L and G components, respectively; same below.) From the frequencies of $f_{2\text{L}}$ and $f_{3\text{L}}$, the asymmetry parameter $\eta \approx 0.088$ is derived at 180 K, indicating a relatively small EFG asymmetry in the disordered state \cite{327and4310-JPSJ2024,4310NMR-NC2026}. The details for estimating $\eta$ and its temperature dependence are presented in Fig.~S2 \cite{SM}. Within the full temperature window of this work, $\eta$ is essentially constant. It seems that both $f_{2\text{G}}$ and $f_{3\text{G}}$ tend to drop below $T_\text{DW}$; however, since we lost NQR signal just below $T_\text{DW}$, it is not clear for us how the EFG changes in the ordered phase.

%Lorentzian fits were performed on the two transition peaks to extract the temperature dependence of the full width at half maximum (FWHM) and the quadrupole frequency in the \textbf{SM} Fig. S3 \cite{SM}.
%Figures \ref{Fig2}(a) and (b) shows the central frequencies of the two transition peaks as a function of temperature.

Figure \ref{Fig2}(c) presents the temperature dependence of $1/T_1T$ at the La(2) site. In the high temperature regime (180 K - 300 K), $1/T_1T$ is nearly constant, reminiscent of a conventional metal behavior without local magnetic moments. Note that for a Fermi-liquid system, $1/T_1T$ is a measure of $N^2(E_{F})$, where $N(E_F)$ is the density of states at the Fermi level \cite{Abragam, Slichter}. %It should be pointed out that the nearly temperature independent $1/T_1T$ behavior at high temperatures contrasts with that of La$_3$Ni$_2$O$_{7}$, where $1/T_1$, rather than $1/T_1T$, levels off at high temperatures \cite{327NQR-CPL2025}, characteristics of spin-dynamics for local moments, suggesting that the electronic correlation is stronger in La$_3$Ni$_2$O$_{7}$. This trend was also supported by recent optical studies \cite{LiuZ-La327Correlation_NC2024, LiuZ-RPCorrelation_PRB2025}.
Upon further cooling, $1/T_1T$ of La$_4$Ni$_3$O$_{10}$ undergoes a pronounced enhancement below 160 K due to the spin fluctuations nearby the SDW order, and then peaks at $\sim 139$ K where the long-range DW transitions occur. The as-defined $T_\text{DW}$ agrees well with that determined by magnetic susceptibility (cf Fig.~S1 \cite{SM}). Below $T_\text{DW}$, as the DW gaps open, presumably, $1/T_{1}T$ should reduce drastically \cite{4310NQR-PRB2026,4310NMR-NC2026}. Unfortunately, since the NQR signals vanish quickly below $T_\text{DW}$ in our experiment, we are unable to see this feature. Notably, above $T_\text{DW}$, a small shoulder is observed. It should be pointed out that the point where $1/T_1T$ starts to upturn coincides with $T^*$ at which the G component in the La(2) NQR spectra appears. Therefore, it is reasonable to attribute the shoulder behavior to spin fluctuations due to the short-range SDW.

%It is worth noting that the L components of both $f_2(T)$ and $f_3(T)$ are essentially featureless across $T^*$, manifesting that the La(2) site is insensitive to charge order in La$_4$Ni$_3$O$_{10}$. However, we will see below that charge dynamics can still be probed by La(2). %In this sense, the observation of the small peak in $1/T_1T$ around $T^*$ suggests that short-range spin order, in addition to short-range charge order, also starts to develop well above $T_\text{DW}$, and this provides additional evidence to the inter-twined DWs in this compound \cite{}.

%\begin{equation}
%\begin{split}
%     \frac{1}{T_{1} T}&=\frac{2 \gamma^{2} k_{\mathrm{B}}}{\gamma_{e}^{2} \hbar^{2}} \sum_{q}\left|A_{q}\right|^{2} \frac{\chi_{\perp}''(\mathbf{q},~\omega)}{\omega}\\
%     &\propto \sum_{q}\left|A_{q}\right|^{2} N(\mathbf{k_{F}}) N(\mathbf{k_{F}}+q)
%\end{split}
%\end{equation}

To further clarify the short-range orders, we analyze La(2) NQR peaks more carefully. The full width at half maximum (FWHM) obtained from the two-component fitting is plotted in Fig.~\ref{Fig2}(e) as a function of temperature. The linewidth for $I>1/2$ nuclei generally receives contributions from both quadrupole ($w^\text{q}$) and magnetic ($w^\text{m}$) interactions, as described by \cite{327NQR-CPL2025,4310NQR-PRB2026,EPJS2010}:
\begin{equation}
   (w_{n})^2 =(w^{\mathrm{q}} _n )^{2}+(w^{\mathrm{m}})^{2}.
   \label{Eq3}
\end{equation}
Theoretically, $w^\text{q}_1 : w^\text{q}_2 : w^\text{q}_3 = 1 : 2 : 3$, whereas $w^{\mathrm{m}}$ has an identical effect on each of the three transition peaks. From the fitting analysis, $w_{2\text{L}}$ and $w_{3\text{L}}$ are found to be nearly temperature-independent, whereas $w_{2\text{G}}$ and $w_{3\text{G}}$ increase markedly with decreasing temperature. To compare their relative evolutions, the FWHM ratio $w_{3\text{G}}/w_{2\text{G}}$ is plotted in the inset to Fig.~\ref{Fig2}(e).
%At $T_{\text{DW}}$, $w_3/w_2$ approaches 1.5, indicating that the quadrupolar contribution dominates the linewidth of the ordered component.
This ratio is about 1.1 near $T^*$, indicating that magnetic broadening is dominant in this regime. It increases gradually upon cooling and attains $\sim 1.5$ near $T_\text{DW}$, revealing a continuous growth of the quadrupolar contribution to the ordered-state linewidth.
%At high temperatures, the absence of local magnetic moments is established from the $1/T_1T$; therefore, the linewidth contribution arises solely from quadrupolar interactions, yielding:
%\begin{equation*}
%\begin{split}
 %   w_{1}:w_{2}:w_{3}= & w_{\mathrm{q}} ^1: w_{\mathrm{q}} ^2 :w_{\mathrm{q}} ^3\\
%    = &1:2:3
%\end{split}
%\end{equation*}
%However, the experimentally observed linewidth ratio $w_3/w_2\approx 1.25<1.5$ significantly deviates from this expected value in Fig. \ref{Fig2}(d) , suggesting that the magnetically broadened component $w_{\mathrm{m}}$ likely originates from itinerant electrons \cite{Co-NMR,La2Ni7-PRB2023}. %Assuming an additional broadening mechanism that affects both transitions equally \cite{CaSb-arxiv2024,LaTiSb-arxiv2024}, analogous to magnetic broadening
%Hence, we obtain:
%\begin{equation}
   %w_{n} =\sqrt{(w_{\mathrm{q}} ^{\pm n/2 \leftrightarrow \pm (n+1)/2} )^{2}+w_{\mathrm{m}}'^{2}   }
%\end{equation}
By combining the linewidths of the $\pm 3/2 \leftrightarrow \pm 5/2$ and $\pm 5/2 \leftrightarrow \pm 7/2$ peaks, we obtain the temperature dependence of  $w^\text{m}_\text{G}$ and $w^\text{q}_\text{G}$ in Fig.~\ref{Fig2}(d). $w^\text{q}_\text{G}$ begins to increase near 150 K, indicating that the spatial distribution of the local EFG starts to broaden. Concurrently, $f_{2\text{G}}$ and $f_{3\text{G}}$ changes abruptly from their L counterparts at this temperature, reflecting a modification of the ensemble-averaged EFG. These features collectively signify that $T^*$ -- which signifies the onset of short-range SDW order -- is also onset of a short-range CDW order. In this sense, we claim here that the La(2) site is also capable of sensing the short-range CDW order that was proposed to appear first in the inner Ni-O layer \cite{4310NMR-NC2026}.

Presumably, $w^\text{m}_\text{G}$ is also anticipated to increase when magnetic order appears, as is the case in La$_3$Ni$_2$O$_7$ \cite{327NQR-CPL2025}. However, to our surprise, a  pronounced decrease rather than increase is observed. Considering the applicability of Eq.~(\ref{Eq3}), when the linewidth contributions from the charge order and the spin order are mutually NOT independent, their interaction may give rise to a coupling term Cov$(\delta \nu^q,\delta \nu^m)$ \cite{2020IJMS-Chen2020}, in which situation Eq.~(\ref{Eq3}) changes into
\begin{equation}
   (w_{n})^2 =(w^{\mathrm{q}} _n)^{2}+(w^{\mathrm{m}})^{2}+2\text{Cov}(\delta \nu^q,\delta \nu^m),
   \label{Eq4}
\end{equation}
where $\delta \nu^q$ is the quadrupolar frequency shift, and $\delta \nu^m$ is the magnetic frequency shift.
Given that the system is undergoing a transition from short-range to long-range order -- a process during which all individual contributions are expected to increase -- one arrives at the conclusion that  Cov$(\delta \nu^q,\delta \nu^m)<0$.
The negative coupling term may originate from antiphase modulation between CDW and SDW order parameters \cite{Tranquada1996,Miao2019}, negative hyperfine coupling constant \cite{Pieper2012}, or negative response coefficient of the local EFG \cite{Kanigel2006}. %\cite{La2Ni7-PRB2023,La139-NMR/NQR-Singer2020}
Most crucially, the phenomena observed in $\mathrm{La}_4\mathrm{Ni}_3\mathrm{O}_{10}$ differ remarkably from those in $\mathrm{La}_3\mathrm{Ni}_2\mathrm{O}_7$ \cite{327NQR-CPL2025}, providing compelling evidence for intertwining between SDW and CDW orders in $\mathrm{La}_4\mathrm{Ni}_3\mathrm{O}_{10}$. %Note that in La$_3$Ni$_2$O$_7$, the SDW and the other DW orders do not appear simultaneously, and they also respond oppositely to pressure: while the SDW order is promoted by pressure, the other DW order is suppressed by pressure \cite{327split-NP2025,327-2025PRB-coexist}.

\subsection{B. $^{139}$La NQR at $p\approx2.3$ GPa}

\begin{figure*}[!ht]
\vspace*{5pt}
\includegraphics[width=17cm]{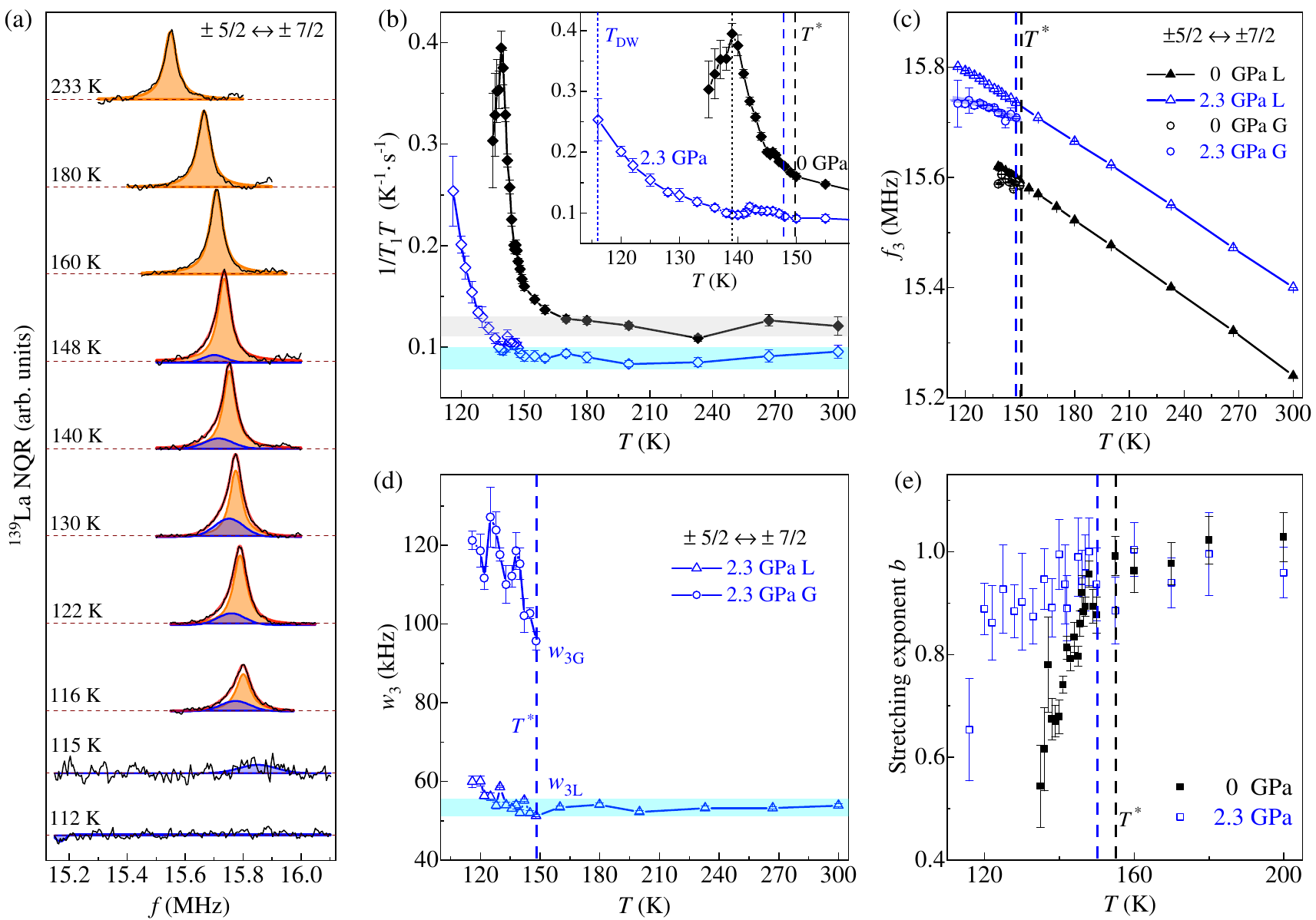}
\vspace*{-5pt}
\caption{(a) La(2) $\pm 5/2 \leftrightarrow \pm 7/2 $ NQR transition peak of La$_4$Ni$_3$O$_{10}$ at 2.3 GPa. The areas colored by orange and blue depict the fittings to Lorentzian and Gaussian functions, respectively. (b) Temperature dependence of $1/T_1T$ at La(2); comparison between ambient pressure (black) and 2.3 GPa (blue). (c) $f_3(T)$. %The inset provides a magnified view of the G component.
(d) $w_3 (T)$. %(e) The extracted $w_\text{q}$ as a function of $T$.
(e) Stretching exponent $b$ obtained from $T_1$ fitting. }
\label{Fig3}
\end{figure*}

The aforementioned difference between La$_4$Ni$_3$O$_{10}$ and La$_3$Ni$_2$O$_{7}$ reminds us to further look into the evolution of the DW orders of La$_4$Ni$_3$O$_{10}$ under pressure. For this purpose, we conducted the same NQR experiments under a hydrostatic pressure of $\sim 2.3$ GPa, and the results are summarized in Fig.~\ref{Fig3}.

Figure \ref{Fig3}(a) displays the La(2) $\pm 5/2 \leftrightarrow \pm 7/2$ transition spectra under 2.3 GPa at selected temperatures (Note \cite{note1}), and Figure \ref{Fig3}(b) shows the temperature dependence of $1/T_1T$ in comparison with the ambient-pressure results. It is clearly seen that now the NQR signal survives until below 116 K, and meanwhile, $1/T_1T$ also appears to peak at around 116 K; in other words, the long-range DW transition now is suppressed to $T_\text{DW}\approx116$ K. Similar to the atmosphere case, a second small peak arising from short-range SDW is also observed above
$T_\text{DW}$ in $1/T_1T$; likewise, its characteristic temperature is defined as $T^*\approx 148$ K, seeing the inset to Fig.~\ref{Fig3}(b). This explicitly implies that the suppression of long-range orders by pressure is much faster than the short-range ones. Another important feature is that the $1/T_1T$ under pressure are consistently lower than those at ambient pressure in the high temperature regime (160 K - 300 K), indicating that pressure reduces the density of states near the Fermi level. This observation is consistent with theoretical predictions \cite{4310NonFermi-PRB2024,4310tend-PRB2024,4310s-wave-PRL2024,4310Hund-SCPMA2025}.

Akin to that at atmosphere, the La(2) NQR resonance peak begins to deviate from the Lorentzian shape near $T^*$, cf Fig.~\ref{Fig3}(a). Therefore, we also decompose the signal into a disordered-state Lorentzian function and an ordered-state Gaussian function. %The extent of suppression is much larger in the short-range order than in the long-range order. Moreover, the $1/T_1T$ under pressure are consistently lower than those at ambient pressure in the high temperature regime (160 K - 300 K), indicating that pressure reduces the density of states (DOS) near the Fermi level. This observation is consistent with theoretical calculations of the pressure induced DOS suppression \cite{4310NonFermi-PRB2024,4310tend-PRB2024,4310s-wave-PRL2024,4310Hund-SCPMA2025}.
The temperature dependencies of $f_3$ for both L and G components are displayed in Fig.~\ref{Fig3}(c). The obtained $f_{3\text{L}}$ is also quasi-linear with $T$, but their values are increased by $\sim 1~\%$ when compared with those at atmosphere. The as-derived $w_3$ for both L and G components are presented in Fig.~\ref{Fig3}(d). Both $f_{3\text{G}}$ and $w_{3\text{G}}$ change abruptly from their L counterparts at $T^*$, indicating that the line broadening below $T^*$ is accompanied with a modification of the EFG, and hence the short-range SDW and CDW are also intertwined under this pressure.

To further elucidate the spin dynamics behavior across the phase transitions, we present in Fig.~\ref{Fig3}(e) the stretching exponent $b$ from the $T_1$ fitting. In general, $b$ quantifies the homogeneity of spin-lattice relaxation process. The necessity of employing this stretching exponent to the fitting is demonstrated in Fig.~S3 \cite{SM}. First of all, a common feature for $p=0$ and 2.3 GPa is that $b$ is close to 1 at high temperature, and starts to deviate obviously right below $T^*$ where short-range DW orders come into being. This suggests that the appearance of short-range DW orders is responsible for the reduction of relaxation homogeneity. Another salient feature is that the decrease of $b$ under pressure is much slower than that at atmosphere.
%higher homogeneity than at ambient pressure¡ªconfirming the excellent hydrostaticity of the applied load. Furthermore, the temperature at which the spin lattice relaxation begins to become inhomogeneous coincides with the onset of short-range order, further confirming the formation of the short-range state \cite{4310NMR-NC2026,Arsenault2018,Baek2017}.

\begin{figure}[!ht]
\vspace*{5pt}
\hspace*{-10pt}
\includegraphics[width=8cm]{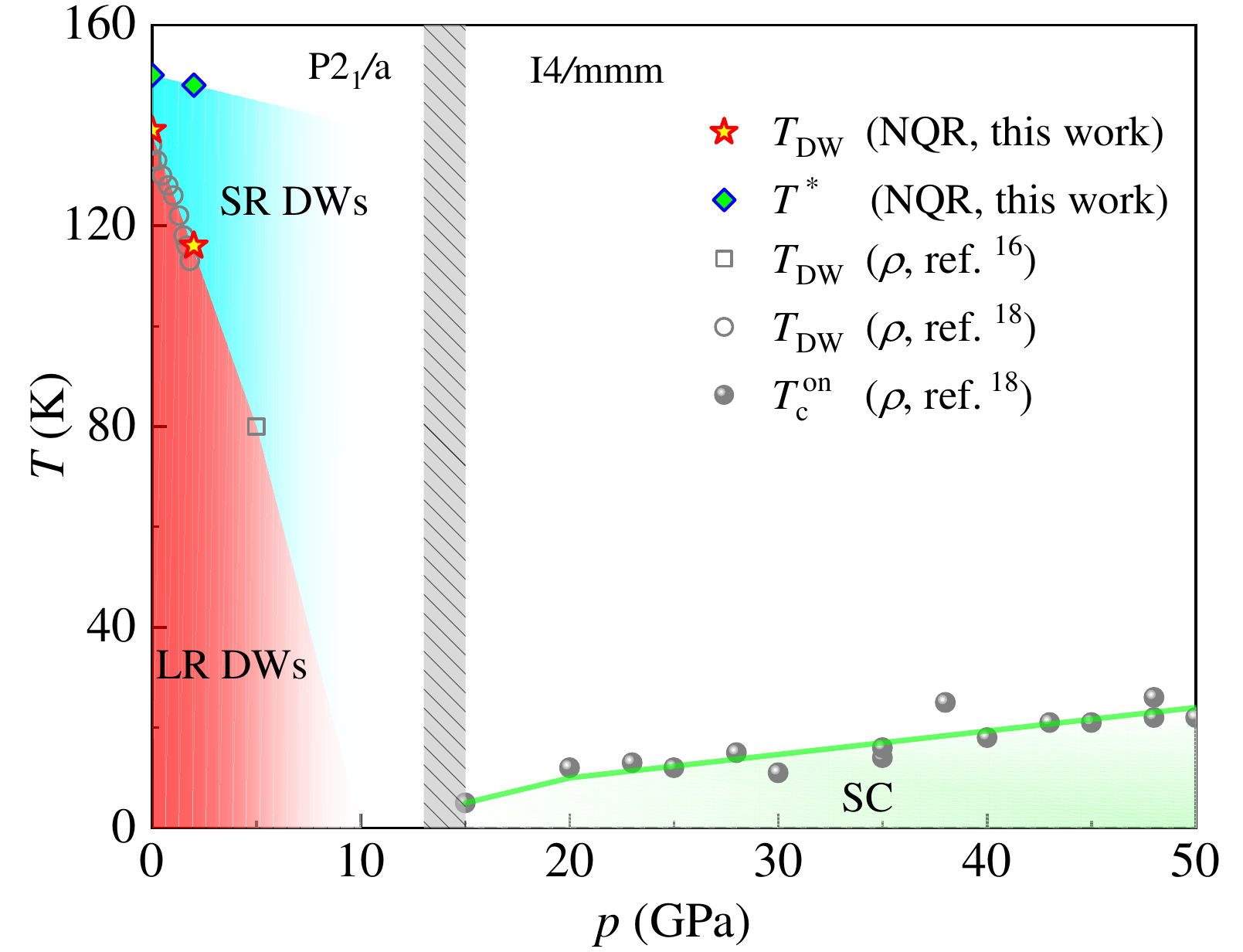}
\vspace*{-5pt}
\caption{Pressure - temperature phase diagram of $\mathrm{La}_4\mathrm{Ni}_3\mathrm{O}_{10}$. In the monoclinic phase (low pressure), short-range DW orders develop upon cooling at $T^*$, followed by the establishment of long-range (LR) DW orders at $T_\text{DW}$. The CDW and SDW orders are intertwined and likely suppressed simultaneously by pressure, up to 2.3 GPa.}
\label{Fig4}
\end{figure}

Based on these results, we construct the pressure - temperature phase diagram,
as shown in Fig.~\ref{Fig4}. The pressure dependent $T_\text{DW}$ derived in this work is in line with the previous transport measurements \cite{4310-Nature2024,4310-PRX2025}. Moreover, our study reveals the emergence of short-range intertwined DW orders prior to the long-range DW transitions. Remarkably, while the short-range DW orders are suppressed at a rate of $1$ K/GPa, the long-range orders exhibit a significantly faster suppression rate of $10$ K/GPa. This leads to a larger window of short-range DW phase under pressure on the phase diagram, which is compatible with the much slower reduction of $b$ as mentioned above. %The pressure-induced suppression rate of this CDW and SDW long-range order is consistent with the transport measurements under pressure, whereas the suppression rate of the short-range order aligns with the results from ultrafast spectroscopy under pressure \cite{4310spectroscopy-NC2025}.

It is worthwhile to compare the phase diagram of La$_4$Ni$_3$O$_{10}$ with that of the bilayer $\mathrm{La}_3\mathrm{Ni}_2\mathrm{O}_{7}$. A major difference is that the SDW order was found to be enhanced by pressure in La$_3$Ni$_2$O$_7$, which not only suggests that the SDW order is decoupled from the lower-temperature DW, but also indicates that SDW does not compete with superconductivity. The enhanced competition between the coexisting intertwined density waves and superconductivity in La$_4$Ni$_3$O$_{10}$ is probably a factor to the reduction of $T_\text{c}$.

Finally, since our experiments only probed the interlayer La(2) sites and thus lacked the information on the La(1) sites, we are unable to confirm the sequential layer effect in the formation of SDW and CDW as was previously reported by other groups \cite{4310NMR-NC2026,4310intertwined-NC2020}. $^{139}$La nuclear magnetic resonance (NMR) experiments under pressure will be needed to complement the present work. Nevertheless, our detailed NQR measurements demonstrate that weak short-range ordered SDW and CDW can be detected at the interlayer La(2) sites.

%Discussion: We analyze the deviation of the FWHM ratio between the $\pm 5/2\leftrightarrow \pm 7/2$ and $\pm 3/2\leftrightarrow \pm 5/2$ transitions from the expected value of 1.5 in the high-temperature regime. Given the Korringa behavior at high temperatures, the absence of local magnetic moments implies that the average static component of the hyperfine coupling is zero; thus, hyperfine interactions are ruled out. Furthermore, dipolar¨Cdipolar interactions contribute only 1\% to the linewidth, suggesting the presence of additional factors. Considering that La(2) resides in the interlayer space of the monoclinic lattice, we speculate that the anomaly arises from either interlayer lattice distortions \cite{4310-crystal-PRB2020,4310Crystal-PRB2018} or vacancies of the apical oxygen atoms \cite{4310-oxygenvacancies-PRB2026,RSC-Adv.2023}.

\section{\Rmnum{4}. Conclusions}

In summary, NQR measurements on the La(2) site of La$_4$Ni$_3$O$_{10}$ reveal successive short-range and long-range intertwined SDW and CDW orders at ambient pressure. Under a hydrostatic pressure up to 2.3 GPa, the intertwined DW orders appear to be suppressed simultaneously. It is also found that the long-range DW orders are suppressed much faster than the short-range orders. Compared to $\mathrm{La}_3\mathrm{Ni}_2\mathrm{O}_{7}$, the intertwined DW orders in $\mathrm{La}_4\mathrm{Ni}_3\mathrm{O}_{10}$ appear to compete more strongly with superconductivity. %thereby suppressing $T_c$.
These findings shed new light to the interplay between density-wave instabilities and superconductivity in RP-phase nickelates.

\section{Acknowledgments}

The authors thank Yaomin Dai, Tao Wu, Meng Wang, Jun Zhao, and Yuefeng Nie for helpful discussions. This work is supported by the National Key R\&D Program of China (2023YFA1609600 and 2022YFA1602602), National Natural Science Foundation of China (U23A20580 and 52588101), and Beijing National Laboratory for Condensed Matter Physics (2024BNLCMPKF004).

\section{Data availability}
All data that support the findings of this study are available from the corresponding authors
upon request.

%\bibliographystyle{apsrev4-2}
%\bibliography{biblio}

\begin{thebibliography}{56}
\bibitem [{not()}]{note1}%
  \BibitemOpen
  \href@noop {} {}\bibinfo {note} {Since there is a large noise near 10.4 MHz
  in our spectrometer, which hindered the measurements of the La(2) $\pm 3/2
  \leftrightarrow \pm 5/2$ transition at 2.3 GPa. Consequently, only the $\pm
  5/2 \leftrightarrow \pm 7/2$ transition was measured.}\BibitemShut {Stop}%
\end{thebibliography}

\newpage

\renewcommand{\thefigure}{S\arabic{figure}}
\renewcommand{\thetable}{S\arabic{table}}
\renewcommand{\theequation}{S\arabic{equation}}
\onecolumngrid

\newpage

\begin{center}
{\bf \large
Supplemental Material:\\
$^{139}$La nuclear quadrupole resonance studies of pressurized La$_4$Ni$_3$O$_{10}$
}
\end{center}

\setcounter{table}{0}
\setcounter{figure}{0}
\setcounter{equation}{0}
\setcounter{page}{1}

\small
\begin{center}
Meng Zhang$^{1}$, Zhuo Wang$^{1}$, Yantao Cao$^{2,3}$, Yang Yuan$^{1}$, Kangjian Luo$^{1}$, Shanxiang Gao$^{1}$, Hanjie Guo$^{3*}$\email{hjguo@sslab.org.cn}, and Yongkang Luo$^{1\dag}$\email{mpzslyk@gmail.com}\\
$^1${\it Wuhan National High Magnetic Field Center and School of Physics, Huazhong University of Science and Technology, Wuhan 430074, China;}\\
$^2${\it Institute of Physics, Chinese Academy of Sciences, Beijing 100190, China; and } \\
$^3${\it Songshan Lake Materials Laboratory, Dongguan, Guangdong 523808, China.} \\

\date{\today}
\end{center}
\normalsize
\vspace*{0pt}

In this \textbf{Supplemental Material} (\textbf{SM}), we provide additional results that further support the discussion and conclusion in the main text, including magnetic susceptibility $\chi(T)$, determination of asymmetry parameter $\eta$, and comparison of $T_1$ fittings with and without stretching exponent $b$. \\

\supplementtableofcontents

\newpage
\SMsection{SM \Rmnum{1}. M\lowercase{agnetic susceptibility}}

\begin{figure}[!ht]
\vspace*{-10pt}
\hspace*{-0pt}
\includegraphics[width=12cm]{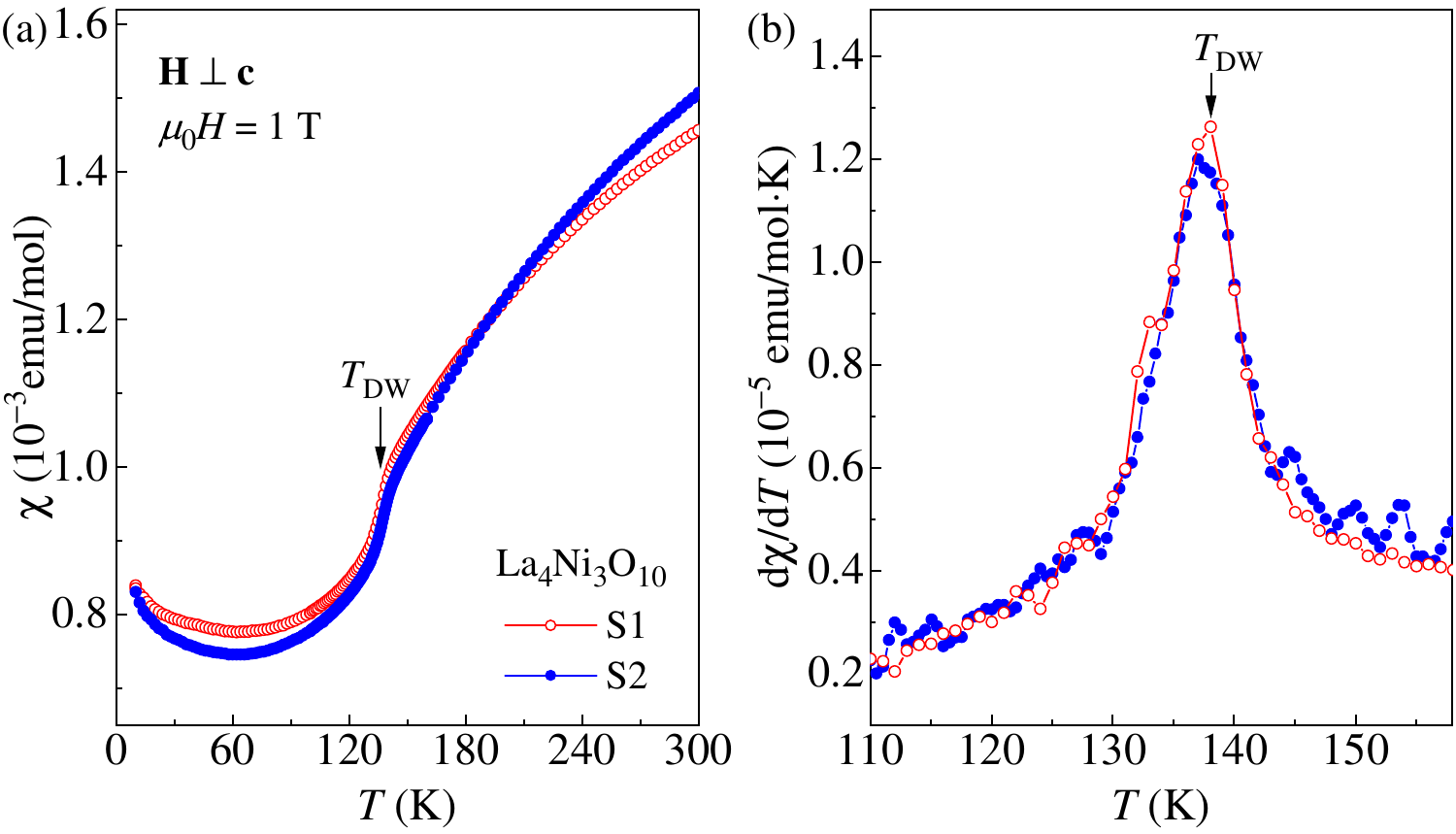}
\vspace*{-0pt}
\caption{Magnetic susceptibility ($\chi$) of La$_4$Ni$_3$O$_{10}$ measured after a zero-field cooling process. Two samples (S1 and S2) were measured. (a) $\chi(T)$; (b) $d\chi/dT$. A density-wave transition (DW) is identified at $T_\text{DW}=139$ K, consistent with literature.}
\label{FigS1}
\end{figure}

\newpage

\SMsection{SM \Rmnum{2}. D\lowercase{etermining the}  EFG \lowercase{asymmetry parameter $\eta$} }

The nuclear quadrupole resonance (NQR) Hamiltonian is given by:
\begin{equation}
    \mathcal{H}_{Q}=\frac{e Q V_{z z}}{4 I(2 I-1)}\left[3 \hat{I}_{z}^{2}-\hat{\mathbf{I}}^{2}+\eta\left(\hat{I}_{x}^{2}-\hat{I}_{y}^{2}\right)\right].
\end{equation}

Diagonalization is achieved through the eigenvector matrix $U$:
\begin{equation}
    \mathcal{H}_{p}=U \mathcal{H}_{Q} U^\dagger.
\end{equation}

\begin{figure*}[!ht]
\vspace*{-0pt}
\includegraphics[width=14cm]{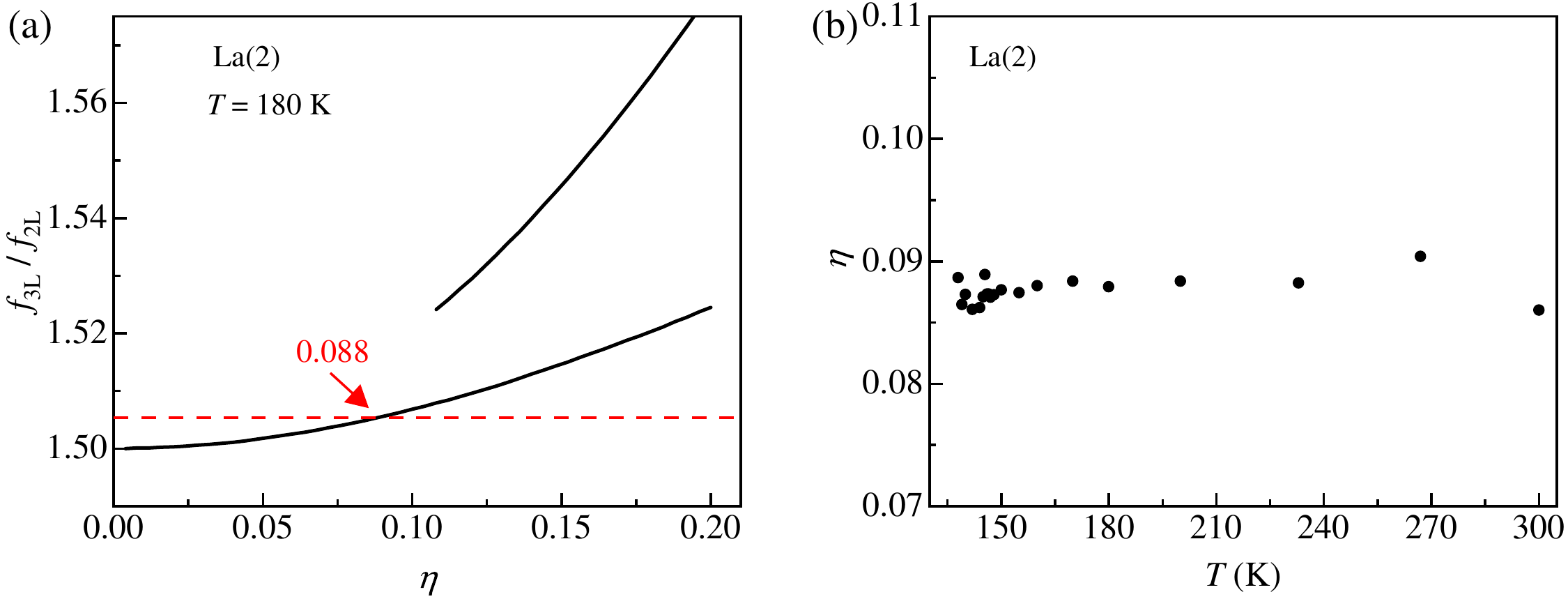}
\vspace*{-0pt}
\caption{ (a) By iterating over different $\eta$ values, the relationship between the  frequency ratio $f_3/f_2$ and $\eta$ is established in the figure. Given that $f_\text{3L}/f_{2L} \approx 1.505$ from experimental, the asymmetry parameter $\eta \approx 0.088$ is obtained at 180 K. (b) Temperature dependence of $\eta$ at the La(2) site of La$_4$Ni$_3$O$_{10}$. }
\label{FigS2}
\end{figure*}

The transition frequencies are derived by the absolute differences between the eigenvalues,
\begin{equation}
    f= |\mathcal{H}_{p}(i,i)- \mathcal{H}_{p}(j,j)|/h.
\end{equation}
$f_2$ and $f_3$ are obtained in this way, and their ratio $f_{3}/f_2$ as a function of $\eta$ can be computed and shown in Fig.~\ref{FigS2}, whose comparison with the experimental yields the best fit $\eta\approx0.088$ at 180 K.

\newpage
\SMsection{SM \Rmnum{3}. F\lowercase{itting of} $T_1$ \lowercase{recovery curves} }

\begin{eqnarray}
\begin{aligned}
M(t)=&M(\infty) \{1-2 F [\frac{2574}{12012}\exp(-(\frac{3t}{T_1})^b)\\
&+\frac{7800}{12012}\exp(-(\frac{10t}{T_1})^b)+\frac{1638}{12012}\exp(-(\frac{21t}{T_1})^b)]\},
\end{aligned}
\end{eqnarray}

\begin{figure*}[!ht]
\vspace*{-10pt}
\includegraphics[width=16cm]{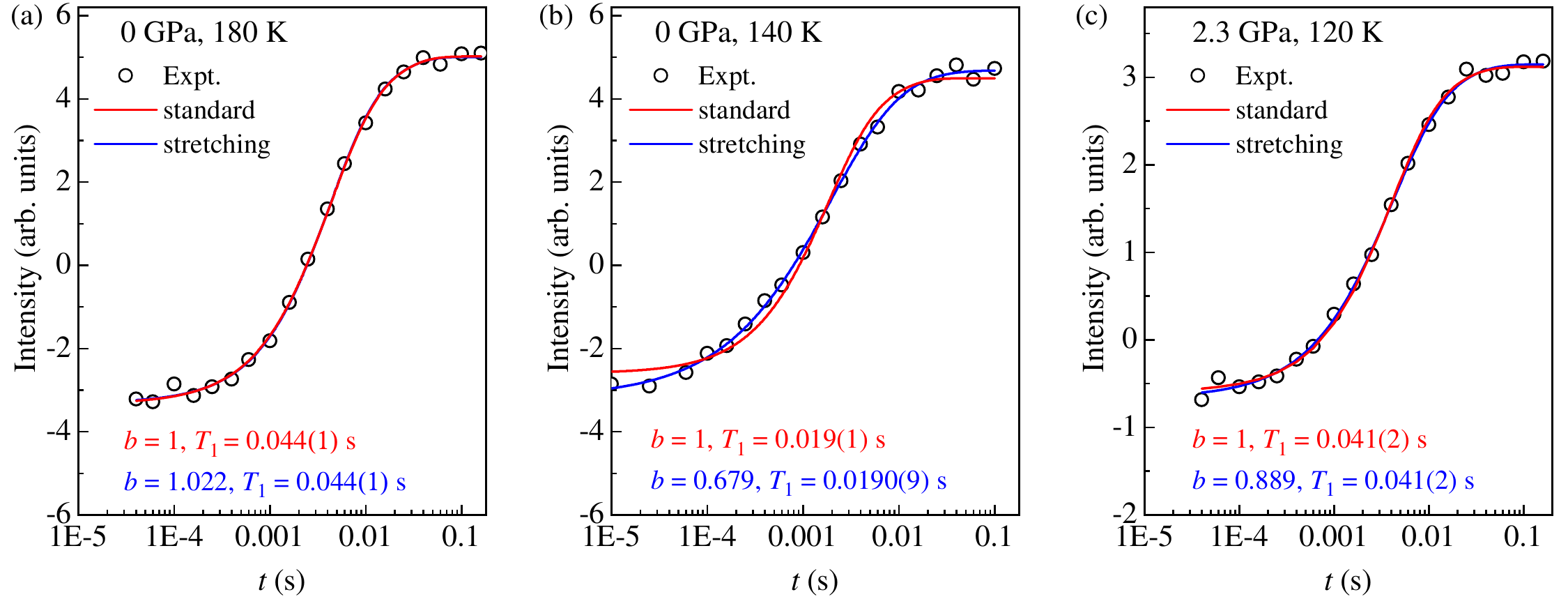}
\vspace*{-5pt}
\caption{ The spin-lattice relaxation time ($T_1$) was obtained by fitting the recovery curve of the $\pm 5/2 \leftrightarrow \pm 7/2$ transition. $b$ is stretching exponent. A smaller b corresponds to a broader $T_1$ distribution, indicative of greater inhomogeneity of the local environment. (a) At 180 K and ambient pressure, both stretched and standard fits yield similar results. (b) At 140 K and ambient pressure, $b=0.679$. (c) At 120 K and 2.3 GPa, $b=0.889$.}
\label{FigS3}
\end{figure*}

%\newpage

%\SMsection{SM \Rmnum{4}. F\lowercase{itting of $^{139}$}L\lowercase{a} NQR \lowercase{peaks}}

%\begin{figure*}[!ht]
%\vspace*{-10pt}
%\includegraphics[width=16cm]{FigS4.pdf}
%\vspace*{-0pt}
%\caption{ Lorentzian fitting of the transition peaks at ambient pressure. (a) Fitting at 160 K, where the peak profile basically follows a Lorentzian shape. (b) Fitting at 146 K; below the short-range density-wave phase transition temperature, it can be observed that the peak profile deviates from the Lorentzian shape. (c) Fitting at 138 K; below the long-range density-wave transition temperature, the peak shape clearly deviates from the Lorentzian lineshape. }
%\label{FigS4}
%\end{figure*}

%\SMsection{SM \Rmnum{5}. F\lowercase{itting of $^{139}$}L\lowercase{a} NQR \lowercase{peaks}}

%\begin{figure*}[!ht]
%\vspace*{-10pt}
%\includegraphics[width=16cm]{FigS5.pdf}
%\vspace*{-0pt}
%\caption{Lorentzian fitting of the transition peaks under pressure. (a) Fitting at 150 K, where the peak profile basically follows a Lorentzian shape. (b) Fitting at 142 K; below the short-range density-wave transition temperature, the peak shape begins to deviate from the Lorentzian lineshape.(c) Fitting at 116 K; below the long-range density-wave transition temperature, the peak shape clearly deviates from the Lorentzian lineshape. }
%\label{FigS5}
%\end{figure*}

\end{document}